\documentclass[10pt]{article}
\usepackage{tabularx}
\usepackage{graphicx}
\usepackage{subcaption}
\usepackage{amsmath}
\usepackage{amssymb}

\usepackage{bbm}
\usepackage{enumitem}

\begin{document}
%
\title{Volunteers in the Smart City: Comparison of Contribution Strategies on Human-Centered Measures}
\author{Stefano~Bennati,~Ivana~Dusparic,~Rhythima~Shinde,~Catholijn~M.~Jonker}%
\date{}




\maketitle


\small\emph{This work has been submitted to the IEEE for possible publication. Copyright may be transferred without notice, after which this version may no longer be accessible.}



\begin{abstract}
  Several smart city services rely on users contribution, e.g., data, which can be costly for the users in terms of privacy. High costs lead to reduced user participation, which undermine the success of smart city technologies.

  This work develops a scenario-independent design principle, based on public good theory, for resource management in smart city applications, where provision of a service depends on contributors and free-riders, which benefit from the service without contributing own resources.

  Following this design principle, different classes of algorithms for resource management are evaluated with respect to human-centered measures, i.e., privacy, fairness and social welfare. Trade-offs that characterize algorithms are discussed across two smart city application scenarios.
  These results might help Smart City application designers to choose a suitable algorithm given a scenario-specific set of requirements, and users to choose a service based on an algorithm that matches their preferences.

\end{abstract}

%

\section{Introduction}

Recent years have seen a substantial increase in active user participation in the smart city, and with it an increase in the resources contributed by the citizens.
Sensor data is one type of user-contributed resource that is at the base of many smart city applications \cite{gaur2015smart}.
Collecting data in a smart city allows to predict the needs of the citizens \cite{al2015applications}, thus enabling the creation of more advanced and more efficient services with a high potential for innovation \cite{hashem2016role}.

User participation is crucial for the success of several smart city applications, but it entails costs that disincentivize users. For example, transmitting privacy-sensitive data in a participatory sensing scenario increases the risk of disclosure and misuse of private information, e.g., discrimination.  Similarly, increasing energy availability in the smart grid scenario, e.g., by postponing the use of appliances, comes with a risk of unfair treatment and disproportinatelly low access to the resource \cite{Bennati2017a}.
In order for the smart city applications to be successful, these costs have to be reduced.
Examples of existing solutions for reducing contribution costs are privacy-enhancing technologies \cite{eckhoff2017privacy} that reduce the risks associated with disclosing information \cite{Finster2015}, and fair resource-management technologies \cite{Tham2014}, which improve the perceived fairness of the system.

Comparing different implementations of these mechanisms is difficult because they are tied to specific assumptions about the scenario and the resource.
To enable a comparison of different scenarios, resources and contribution strategies, this paper introduces a design principle for developing smart city algorithms, which allows the evaluation of privacy-enhancing technologies in a scenario-independent approach.
Several smart city services are produced from user-contributed data. Proposed approach is based on the theory of public goods and voluntary contribution games \cite{Bagnoli1989}, as it is an approach well suited for modeling scenarios where a common resource, i.e. the smart-city service, depends on the action of a population \cite{Wolsink2012}.
The orchestration of demand and offer of some resource is driven by contribution strategies, algorithms that determine which users should contribute to the system at each point in time, based on the state of the system and a set of system requirements.
By acting on the decision of whether to participate, a contribution strategy is independent of the characteristics of the resource, hence it can be combined with existing mechanism that work at the content level, e.g., privacy-preserving algorithms.
Additionally, it makes data sharing across smart city applications easier, thanks to the uniform way of dealing with different kinds of data.

Following this design principle, a simulation framework is developed \cite{code} that allows a comparison of multiple algorithms for resource contribution, to address the research question: how do different contribution strategies compare in terms of privacy, fairness and social welfare?

The first contribution of this paper is to introduce the design principle and verify its applicability to two different application scenarios, i.e., traffic congestion information and charging of electric vehicles, by means of real-world datasets. Similar validation in different scenarios can provide useful insights for the design of other resource-based smart city applications.

The second contribution is to showcase the generality of the framework by implementing two classes of algorithms -- centralized optimization and localized reinforcement learning in two flavors, with and without contextual awareness -- in these two smart city scenarios and evaluating them in terms of trade-offs between system efficiency, user privacy or resource distribution fairness, as opposed to optimizing them towards a single measure.

The results presented show that the same trade-offs between algorithms repeat across the application scenarios. Specifically, a localized solution, which works only on local data, is found to deliver higher privacy and equality than a centralized solution.
Centralized optimization offers instead higher efficiency thanks to the global knowledge of data.
Localized algorithms with and without contextual knowledge are evaluated, and those considering the context in the decision-making are found to deliver higher fairness.

This work can be of interest to designers of smart city applications and services that look for a guideline for the choice of algorithm, given specific scenario priorities in terms of different measures. Another target audience is that of citizens that are concerned with the risk of abuse of their data, and particularly privacy-aware prosumers.
The results help quantify different service implementations along measures such as privacy and fairness, thus help the citizens choose a service provider that best matches their preferences, reducing the concerns about the system and fostering user participation.

The rest of the paper is organized as follows: Section \ref{sec:model} introduces the design principle and how it can be applied to two smart city scenarios, Section \ref{sec:results} describes and discusses the main results obtained from simulating voluntary contributions in different scenarios, and Section \ref{sec:conclusions} concludes the paper discussing possible avenues for future work and giving general recommendations about the choice of contribution strategies.

\section{Model}
\label{sec:model}

This section introduces the design principle, based on the theory of public goods and Voluntary Contribution Games  \cite{Bagnoli1989}, and it describes how it can be applied to the smart city scenarios of participatory sensing, with the example of traffic congestion \cite{Duan2014}, and smart grids, with the example of electric vehicle charging \cite{wang2013survey}.
Voluntary contribution games involve the need by a number of users to coordinate the provision of some public good, e.g., financing a playground with private donations.
Table \ref{tbl:notation} illustrates the notation of the model, symbols are listed in order of appearance.

\begin{table}[!htb]
\centering
{\footnotesize
\begin{tabularx}{\columnwidth}{@{}lX@{}}
Math symbol & Description\\
  \hline
    $S$ & The service provider \\
  $V=\{ 1, \ldots , n\}$ & The set of $n$ users \\
  $T \in \mathbb{N}_{>0}$ & The number of rounds in the simulation \\
  $r_i^t$ & The resource produced by user $i$ at time $t$ \\
  $v_i^t \in \mathbb{R}$ & The value associated to disclosing $r_i^t$ \\
  $c_i^t \in \mathbb{R}$ & The cost associated to transmitting $r_i^t$ \\
  $p_i^t \in \mathbb{R}$ & The privacy leaked when disclosing $r_i^t$ \\
  $\mathcal{A} = \{D,C\}$ & The action set \\
  $a_i^t \in \mathcal{A} $ & The action of user $i$ at time $t$ \\
  $A^t_{+} \subseteq V$ & The set of contributors at time $t$ \\
  $q^t \in \mathbb{R}$ & The service quality at time $t$\\
  $f: V \rightarrow \mathbb{R}$ & The quality function \\
  $\tau^t \in \mathbb{R}$ & The quality requirement for time $t$\\
  $S : \mathbb{R}\times\mathbb{R} \rightarrow \mathcal{R}$ & The success function \\
  $G^t \in \mathbb{R}$ & The payoff for a successful round at time $t$\\
  $G : \mathbb{R} \times \mathbb{R} \rightarrow \mathbb{R}$ & The payoff function for successful rounds \\
  $B^t \in \mathbb{R}$ & The payoff for an unsuccessful round at time $t$\\
  $B : \mathbb{R} \times \mathbb{R} \rightarrow \mathbb{R}$ & The payoff function for unsuccessful rounds \\
  $U_i^t$ & The utility of user $i$ at time $t$\\
  $Q$ & The total quality of service after $T$ rounds \\
  \hline
            & \textbf{Scenario: smart grid} \\
              $\pi^t_i \in \mathbb{R}$ & The energy production of household $i$ at time $t$\\
  $\beta^t_i \in \mathbb{R}$ & The baseline consumption of household $i$ at time $t$\\
  $\sigma^t \in \mathbb{R}$ & The energy surplus at time $t$\\
\hline
\end{tabularx}
}
\caption{Mathematical notation, in order of appearance.}\label{tbl:notation}
\end{table}

$V=\{ 1, \ldots , n\}$ is the set of users and $S$ is a service provider.
In each round $t \leq T$ each user produces a resource $r_i^t$ with value $v_i^t$.
Users perform an action $a_i^t \in \mathcal{A}=\{C,D\}$: $C$ corresponds to contributing the resource and $D$ to opt out from contribution.
The set of contributors at time $t$ is denoted as $A^t_{+}=\{ i \in V: a_i^t=C\}$.

Contribution is costly, thus contributors pay a cost $c_i^t$ that depends on the characteristics of resource and communication medium.
Contribution might also entail a privacy cost $p_i^t$, which models the risks of revealing private information to third parties.

The service provider determines the service quality $q^t=f(A^t_{+})=\sum_{i \in A^t_{+}} v_i^t$ based on all contributions received.
A quality requirement $\tau^t$, either global or per user, is generated at every timestep. Not all users are required to contribute in order to satisfy the requirement $\tau^t\le\sum_i v_i^t$.
A round is successful, i.e. the service can be provided, if the quality is higher than the threshold: the success function is defined as $S(q^t,\tau^t)=\{ G^t ~\mbox{if}~ q^t \ge \tau^t ~\mbox{else}~ B^t \}$.
Each user gets the same positive payoff $G^t=G(\tau^t,q^t)$ from accessing the service, including those who did not contribute.
If the quality threshold is not met, the service cannot be provided and every agent receives a large negative payoff $B(\tau^t,q^t)>0$.
Given that payoffs are distributed equally, there is no incentive for users to contribute unilaterally: the public goods theory predicts the existence of an equilibrium where nobody contributes.

The individual goal of the users is to maximize their individual utility over $T$ rounds (see Table \ref{fig:game}): $U_i^t= G(\tau^t,q^t)\mathbbm{1}_{q^t \ge \tau^t}-B(\tau^t,q^t)\mathbbm{1}_{q^t < \tau^t}-c_i^t\mathbbm{1}_{a_i^t}$, where $\mathbbm{1}_{X}$ is the indicator function (equals 1 if event X is true).
The social goal, i.e. the goal of the service provider, is to maximize the total quality of the service over $T$ rounds $Q=\max \sum_{t=1}^T f^t(A^t_{+})$.

The model relies on the following assumptions:
\begin{itemize}
\item Users contribute their resources to a central entity, which uses them to provide a service.
\item Contributions cannot be doctored, e.g., to reduce the cost.
\item Users are not allowed to communicate with each other; this is generally the case for simple devices such as smart meters \cite{fadlullah2011toward}.
\item The system implements a privacy-preserving resource-management algorithm that optimizes the use of the resource. 
\end{itemize}

\subsection{Applicability to real-world scenarios}
\label{sec:scenarios}
This section describes how the scenarios of smart and participatory sensing can be modelled using the proposed design principle.
\subsubsection{Smart grid}
Electric vehicles (EVs), associated to households connected to the smart grid, are required to be periodically recharged.
The \emph{public good} consists of the total energy surplus $\sigma^t=\pi^t-\beta^t$, which is available to all households for charging the EVs, its value is computed from the current production of renewable energy $\pi^t=\sum_i \pi^t_i$, and the current network load $\beta^t=\sum_i \beta^t_i$, obtained from EirGrid data \cite{eirgrid}. 
Contribution to the public good is then defined as opting out from consumption, i.e. renouncing to charge the EV of a charge $v^t_i$ that might depend on contextual variables such as the current charge level, the availability of a charging station, and the current energy surplus.
Contribution entails a comfort cost that is assumed to be proportional to the corresponding value, as the utility of an EV depends on it being charged.
\emph{Values} are determined in the experiments by sampling a uniform probability distribution, and \emph{costs} are determined by sampling a normal distribution, centered around the respective value.
\subsubsection{Participatory sensing}
Traffic congestion is computed by aggregating speeds and locations reported by cars using a participatory sensing approach \cite{Brown2013}.
The value $v^t_i$ of individual measurements reflects the novelty of the information, which might depend on the actions of other users \cite{Vuran2004}, e.g., duplicated information due to local correlation in measurements.
In the specific case, measurements cannot be linked with one another as the dataset lacks GPS coordinates \cite{nrel}, therefore the \emph{value} of novelty is approximated with the change of speed, as a sudden change in speed is considered more informative than keeping a constant speed.
Contributing data comes at a transmission cost $c_i^t$ -- that might depend on the characteristics of the message, e.g., size, and of the medium, e.g., congestion, power consumption -- and a privacy cost $p^t_i$ -- that might depend on disclosing private information, such as location and speed \cite{Tsoukaneri2016}.
\emph{Costs} are determined as the distance to the only points of interests known in the data: the origin and destination of a trip.
The \emph{public good} is created if the sum of individual contributions is higher than a certain threshold $\tau$, which is set to 80\% of the size of the population, which means that the service is successfully generated if at least 80\% of the users contribute a value of 1.

\begin{table*}[t]
  \centering{
\begin{tabular}{|c|c|c|c|}
  $q^t_{-i}$ & $q^t_{-i}+v_i^t < \tau^t$ & $\tau^t-v_i^t \le q^t_{-i} < \tau^t$ & $\tau^t \le q^t_{-i}$\\
  outcome & failure & depends on $a_i$ & success \\
\hline
$U_i^t(a_i=C)$ & $-B(\tau^{\text{t}},q^{\text{t}})-c_{\text{i}}^{\text{t}}$ & $G(\tau^{\text{t}},q^{\text{t}})-c_{\text{i}}^{\text{t}}$ & $G(\tau^{\text{t}},q^{\text{t}})-c_{\text{i}}^{\text{t}}$\\
$U_i^t(a_i=D)$ & $-B(\tau^{\text{t}},q^{\text{t}})$ & $-B(\tau^{\text{t}},q^{\text{t}})$ & $G(\tau^{\text{t}},q^{\text{t}})$\\
\end{tabular}
\subcaption{Definition of utilities for agent $i$. Let $q_{-i}^t=\sum_{j \in A^t_{+} \setminus \{i\}} v_j^t=q^t-v_i^t$ be the total contributions, excluding agent $i$, and $\tau^t$ be the global requirement.
  This game qualifies as a threshold public goods game if $G(\tau^t,q^t)+B(\tau^t,q^t) > c_i^t$, which is always verified for large enough values of $G$ or $B$.}\label{fig:game}
\begin{tabular}{lll}
&Measure & Definition\\
  \hline
  a) &Success & $\Sigma^t=\min(1,q^t/\tau^t)$\\
  b) &Efficiency & $E^t=\tau^t / q^t ~\mbox{if}~ \tau^t \le q^t ~\mbox{else}~ 0$\\
  c) &Social welfare & $W^t= \frac{i}{T} \sum_{t=1}^T U_i^t$\\
  d) &Privacy & $P^t=1-|A^t_{+}| / n$ \\
  e) &Fairness & $F^t=\frac{1}{n} \left ( n+1-2 \left (  \frac{\sum_{i=1}^n (n+1-i)y_i}{\sum_{i=1}^n y_i} \right) \right ) ~where~ y_i=v_i^t$\\
  f) &Fairness, over time & $G=\frac{1}{n} \left ( n+1-2 \left (  \frac{\sum_{i=1}^n (n+1-i)y_i}{\sum_{i=1}^n y_i} \right) \right ) ~where~ y_i=\sum_{t=1}^T v_i^t$\\
\end{tabular}
\subcaption{Measures used during evaluation.}\label{fig:measures}}
\caption{Analytical definition of criteria of performance.}
\end{table*}

\subsection{Measures}
\label{sec:measures}
The performance of different contribution strategies are quantified with the following measures (see Table \ref{fig:measures}):
\begin{enumerate}[label=\alph*)]
\item Success rate: The fraction of the threshold that has been covered by contributions, or 1 if the total contribution is higher than the threshold.
\item Efficiency: The ratio between the requirement and the sum of contributions. Efficiency is 1 if the sum of contributions is equivalent to the threshold e.g., all agents contribute 1, it is lower than 1 if the sum of contributions is larger than the threshold.
\item Social welfare: The average reward, sum of a constant benefit and a constant negative cost (only for contributors).
\item Privacy: the fraction of agents that did not disclose private information during the current timestep.
\item Fairness: The Gini coefficient represents the fairness of the current round of contributions (0 is total equality). It is computed for each timestep, as the Gini coefficient of the values that agents contributed in that timestep, and aggregated across time and repetitions.
\item Fairness over time: It represents the fairness of the history of contributions. It measures the Gini coefficient computed across agents on the individual histories of contribution, from the start of the simulation to the current timestep.
\end{enumerate}
The definition of privacy can be made scenario-specific by adopting an appropriate privacy measure, e.g., K-anonymity, differential privacy \cite{Damiani2014}.

\subsection{Algorithms}
\label{sec:algorithms}

This section describes the contribution strategy algorithms chosen for the analysis in our framework.
The choice favored well-established and general-purpose algorithms as opposed to algorithms with state-of-the-art efficiency, as the goal is to highlight trade-offs between measures over different scenarios.
The review and comparison of scenario-specific algorithms is not aligned to the goals of this paper and is therefore out of scope.

\subsubsection{Centralized algorithms}
Centralized or top-down algorithms rely on a central optimizer that satisfies the public good while minimizing the cost of contribution.
This problem can be modeled as the well-know Knapsack problem: $\begin{cases} \mbox{minimize} & \sum_{i=1}^{n}c_i^t*a_i^t \\ \mbox{subject to} & \sum_{i=1}^{n} v_i^t*a_i^t > \tau^t \end{cases}$,
where the weight of items is given by the contribution value and the value of each item is defined as the inverse of the cost.
We chose a customized ``fully polynomial time approximation scheme'' (FPTAS) that reaches the knapsack constraints from above, instead of from below, such that the threshold can be met.

\subsubsection{Localized algorithms}

Decentralized algorithms distribute decision making at the local level and allow communication between agents for coordination \cite{Kennedy1995} or learning \cite{Dusparic2017}.
In this paper we focus on \emph{localized} algorithms, a type of decentralized algorithms that operate only on local knowledge, without assuming the availability of special hardware for communication \cite{fadlullah2011toward}.
Aspiration learning is a learning algorithm that is specifically tailored for coordination games \cite{Chasparis2010}; agents contribute based on their ``satisfaction value'', which depends on their previous experiences, but does not consider the current context, e.g., current value or cost.
Q-Learning is a model-free unsupervised reinforcement learning approach that considers both the history and the current context in the decision \cite{sutton1998reinforcement}.

A disadvantage of reinforcement learning is its sensitivity to initial conditions, for example a multi-agent learning process might converge to an inefficient equilibrium where nobody contributes.
In order to make this outcome less likely, agents are pre-trained to prefer contribution in order to bias the initial exploration period.
Pre-training is a reasonable solution as it can be performed during device manufacturing and its effect on the behavior of agents fades off quickly as agents start learning.

\section{Results and Analysis}
\label{sec:results}
This section presents the results of computational experiments. Trade-offs are evaluated between the three contribution strategies presented in section \ref{sec:algorithms} and two baselines: a baseline in which users contribute ``fully'', i.e. always, and a ``random'' baseline in which users have 50\% probability of contributing at each round \footnote{50\% chance of contribution does not imply that 50\% chance of success, because each contribution has an average value greater than 1.}.

All experiments are performed in a simulation framework developed in Python and run on ETH’s cluster “Euler”. Results presented in this paper represent the average state of 20 simulations after 5000 timesteps, and error bars represent the confidence intervals at 95\%.

The contribution strategies are tested with real-world datasets in two smart city applications, using cost, value and public good values as described in Section \ref{sec:scenarios}.
The traffic dataset contains mobility traces of private cars obtained from U.S. Department of Energy National Renewable Energy Laboratory\cite{nrel}. The electric vehicle (EV) dataset contains residential consumption data obtained from Irish Smart Meter trial and renewable energy production data from Irish elecriticy grid operator EirGrid \cite{eirgrid}.

Full results are shown in Figure 1 and present the comparisons of the contribution strategies by the six measures discussed in Section \ref{sec:measures}.
Most measures do not show a dependency on the size of the population, i.e., the number of sensors, because the value of the public good threshold is chosen to be proportional to the size of the population.
If the threshold would be constant, an increase in the number of sensors, i.e., potential contributors, would make the creation of the public good easier, hence affect all measures.

\begin{figure*}[!p]
  \centering
    \begin{minipage}[b]{0.9\linewidth}
      \includegraphics[width=\linewidth]{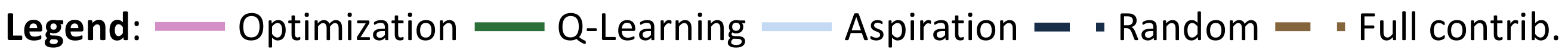}
      \vspace{0pt}
\end{minipage}
\begin{minipage}[b]{\linewidth}
  \begin{minipage}[b]{.33\linewidth}
    \includegraphics[width=\linewidth]{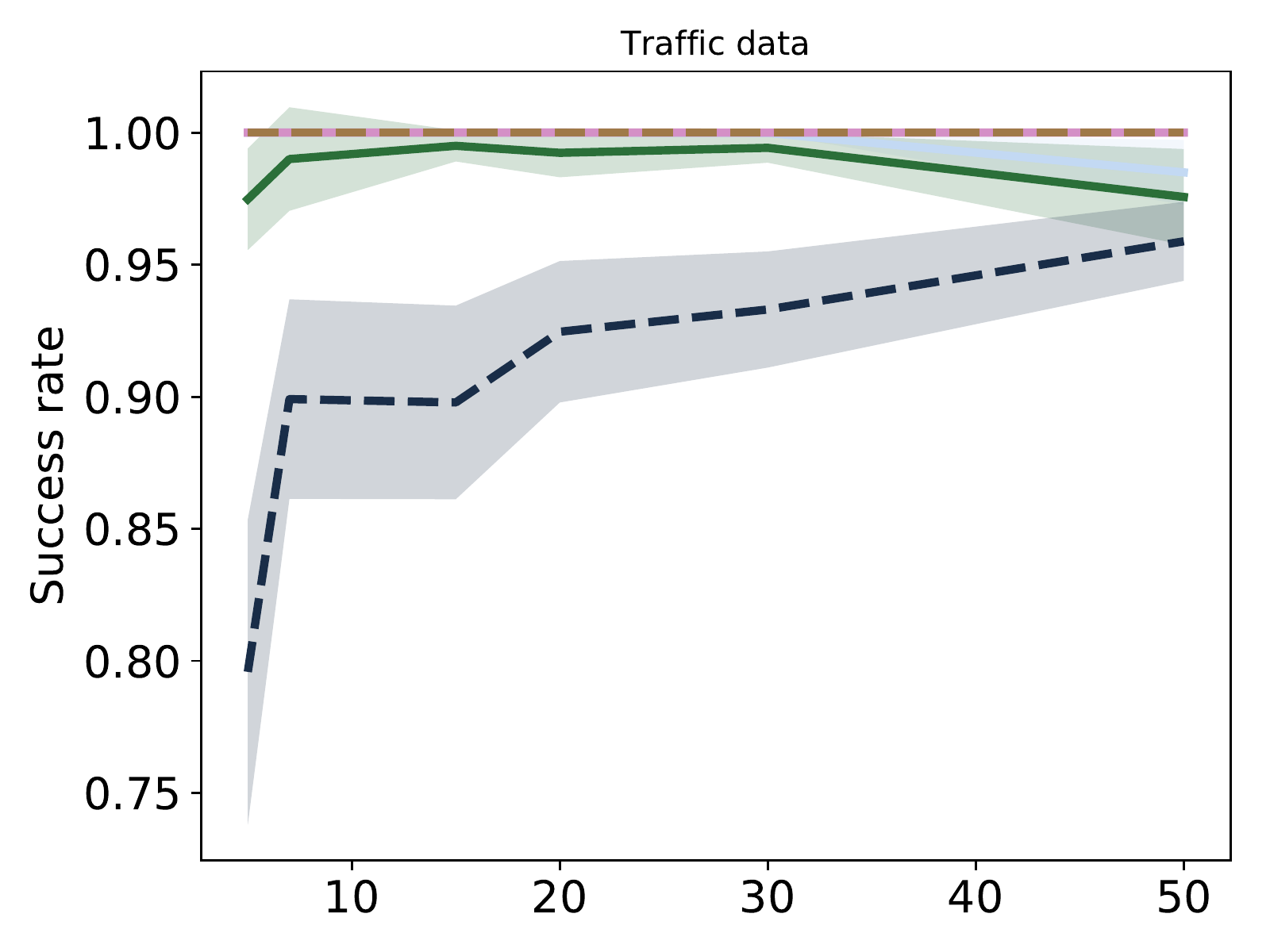}
    \includegraphics[width=\linewidth]{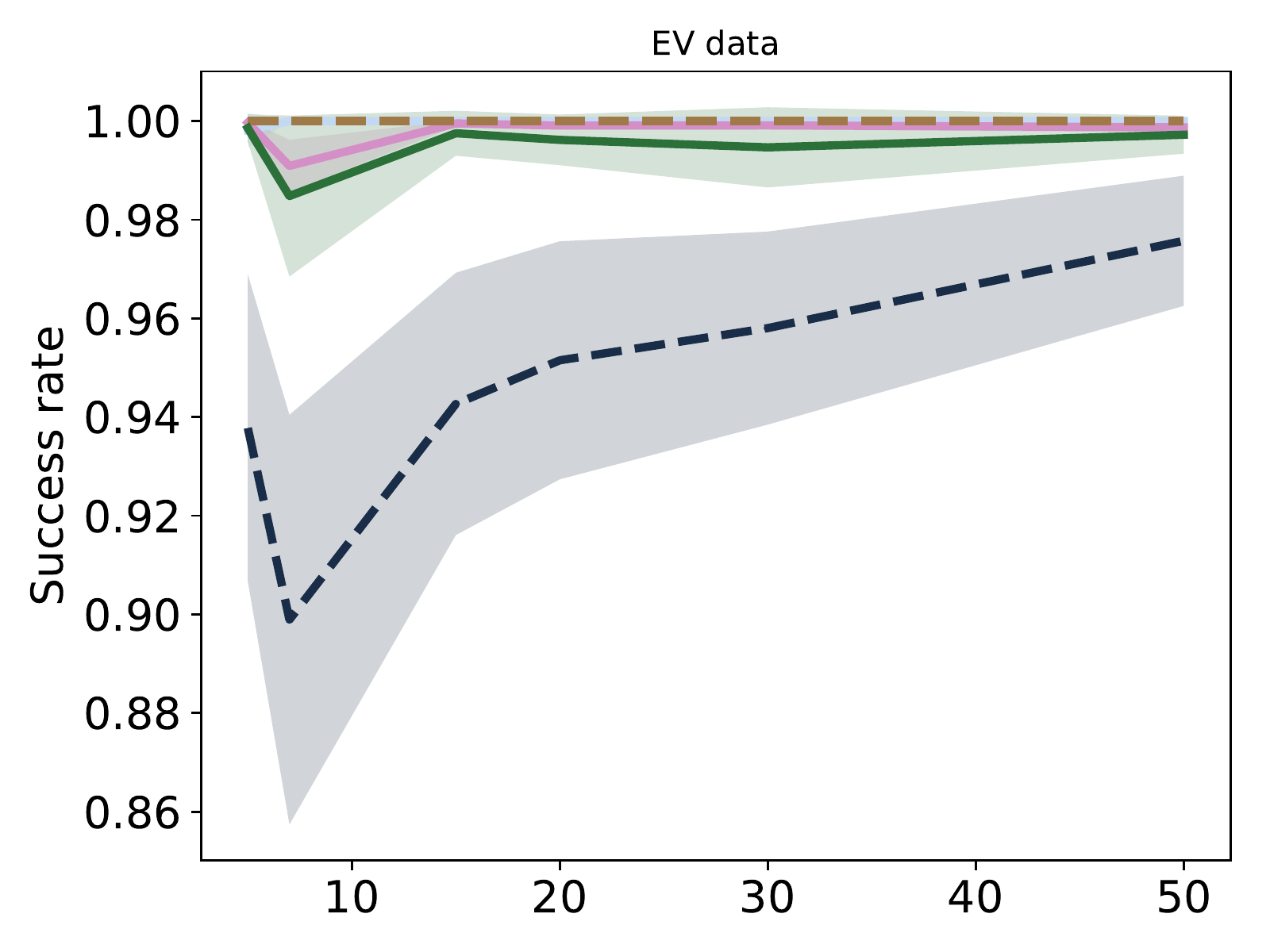}
    \subcaption{Ratio of success in service creation.}\label{fig:succ}
  \end{minipage}%
  \begin{minipage}[b]{.33\linewidth}
      \includegraphics[width=\linewidth]{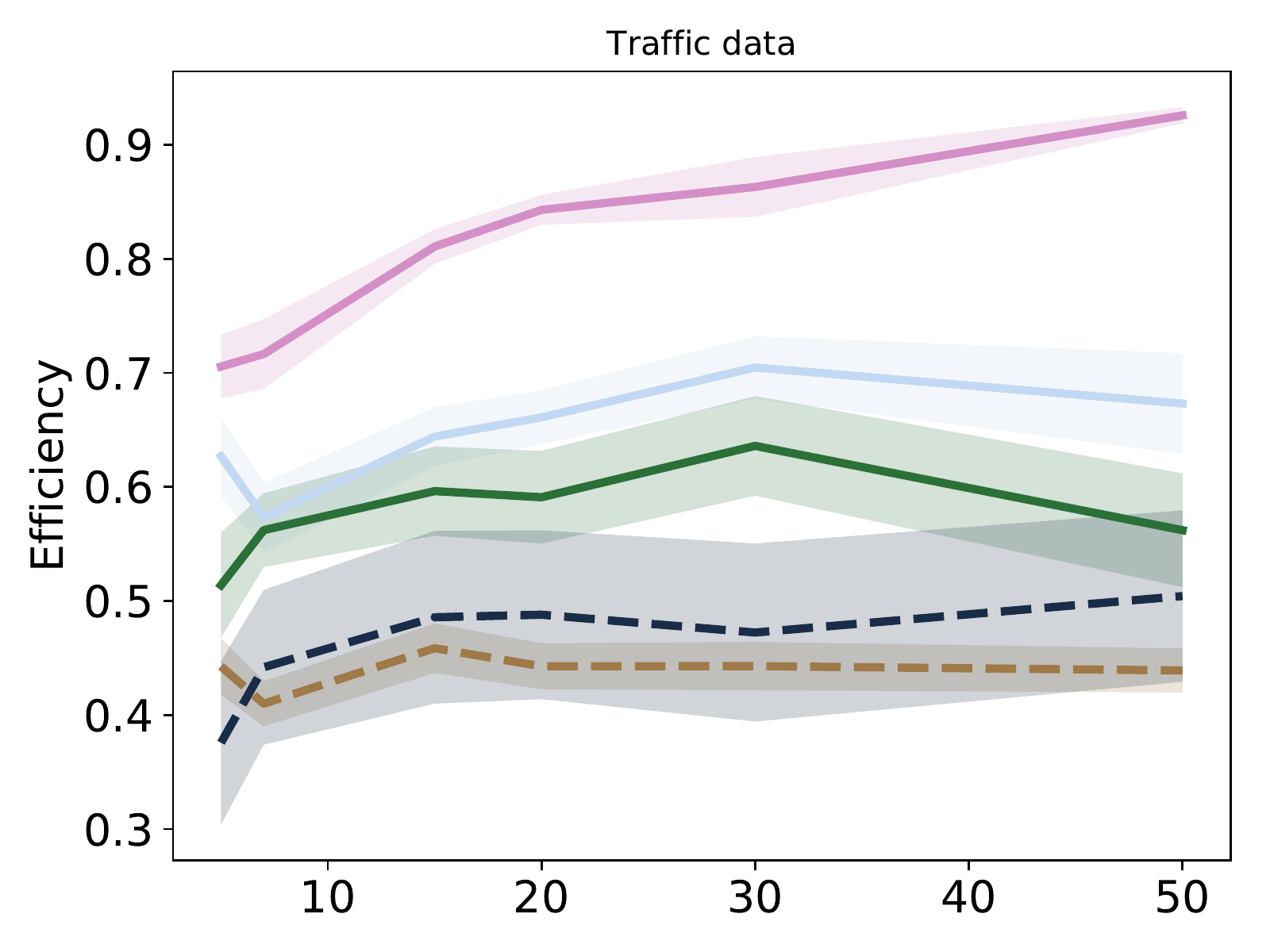}
      \includegraphics[width=\linewidth]{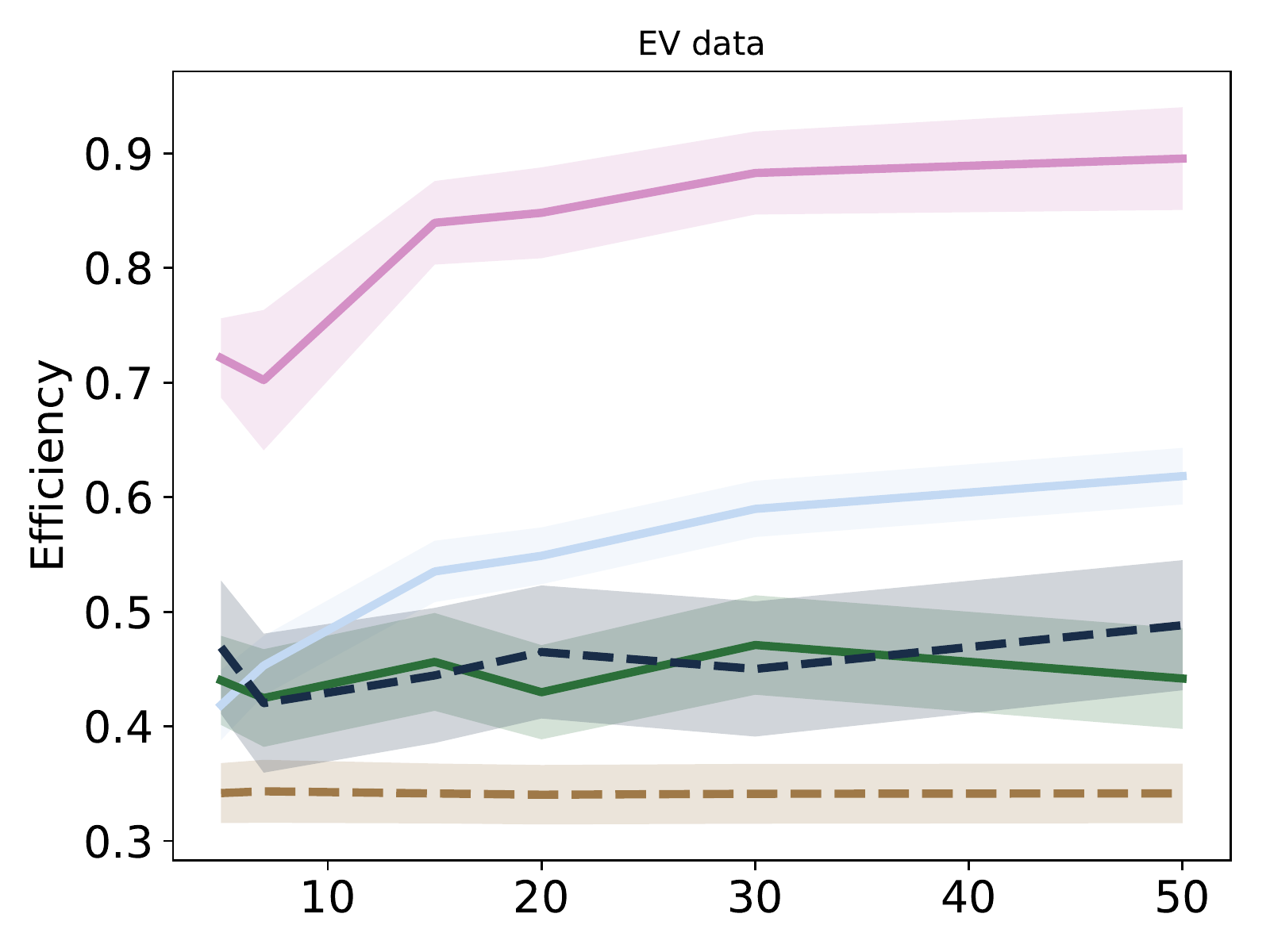}
      \subcaption{Efficiency of total contributions.}\label{fig:eff}
  \end{minipage}%
  \begin{minipage}[b]{.33\linewidth}
      \includegraphics[width=\linewidth]{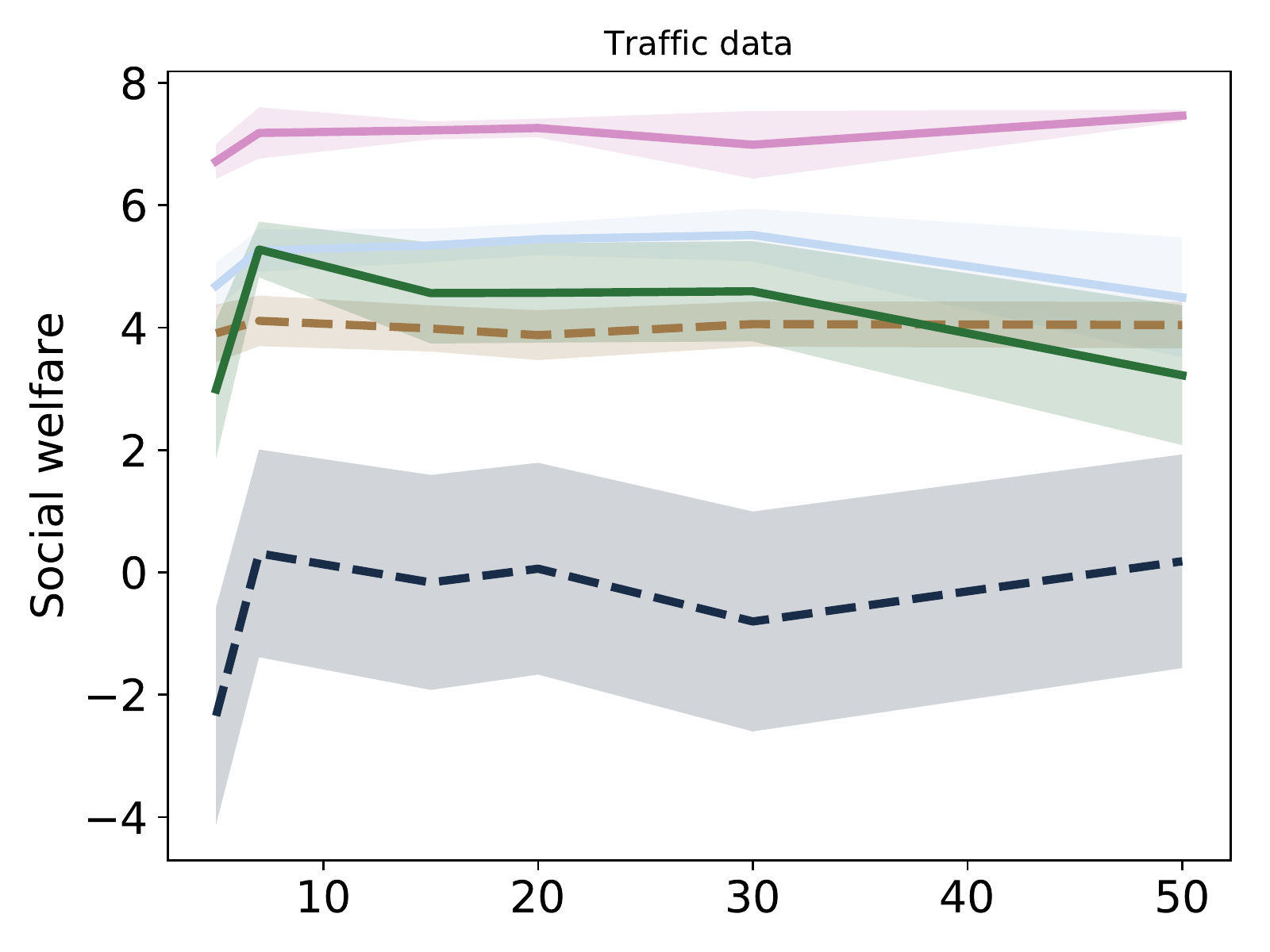}
      \includegraphics[width=\linewidth]{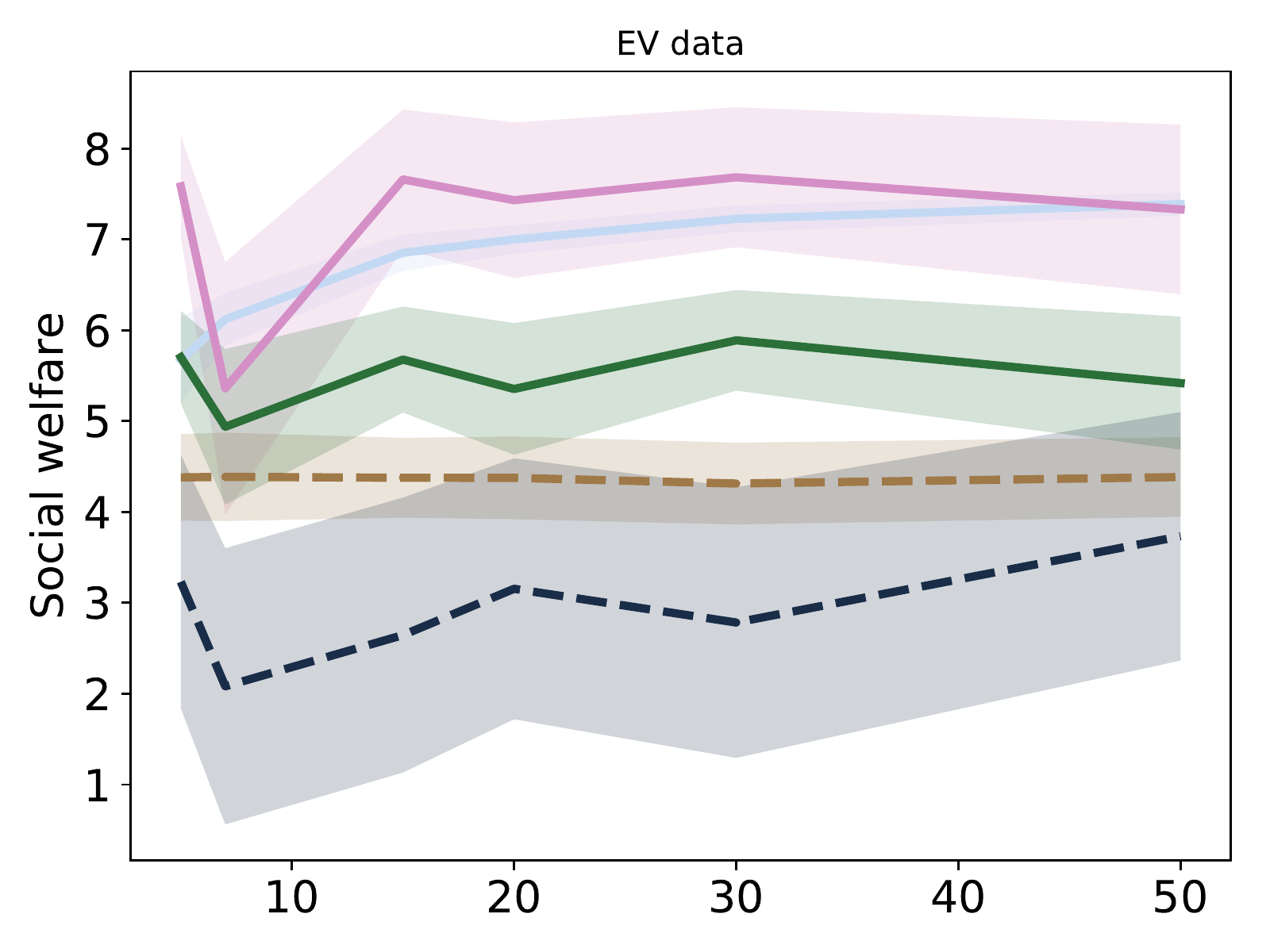}
        \subcaption{Social welfare of the population.}\label{fig:welfare}
  \end{minipage}
  \begin{minipage}[b]{.33\linewidth}
      \includegraphics[width=\linewidth]{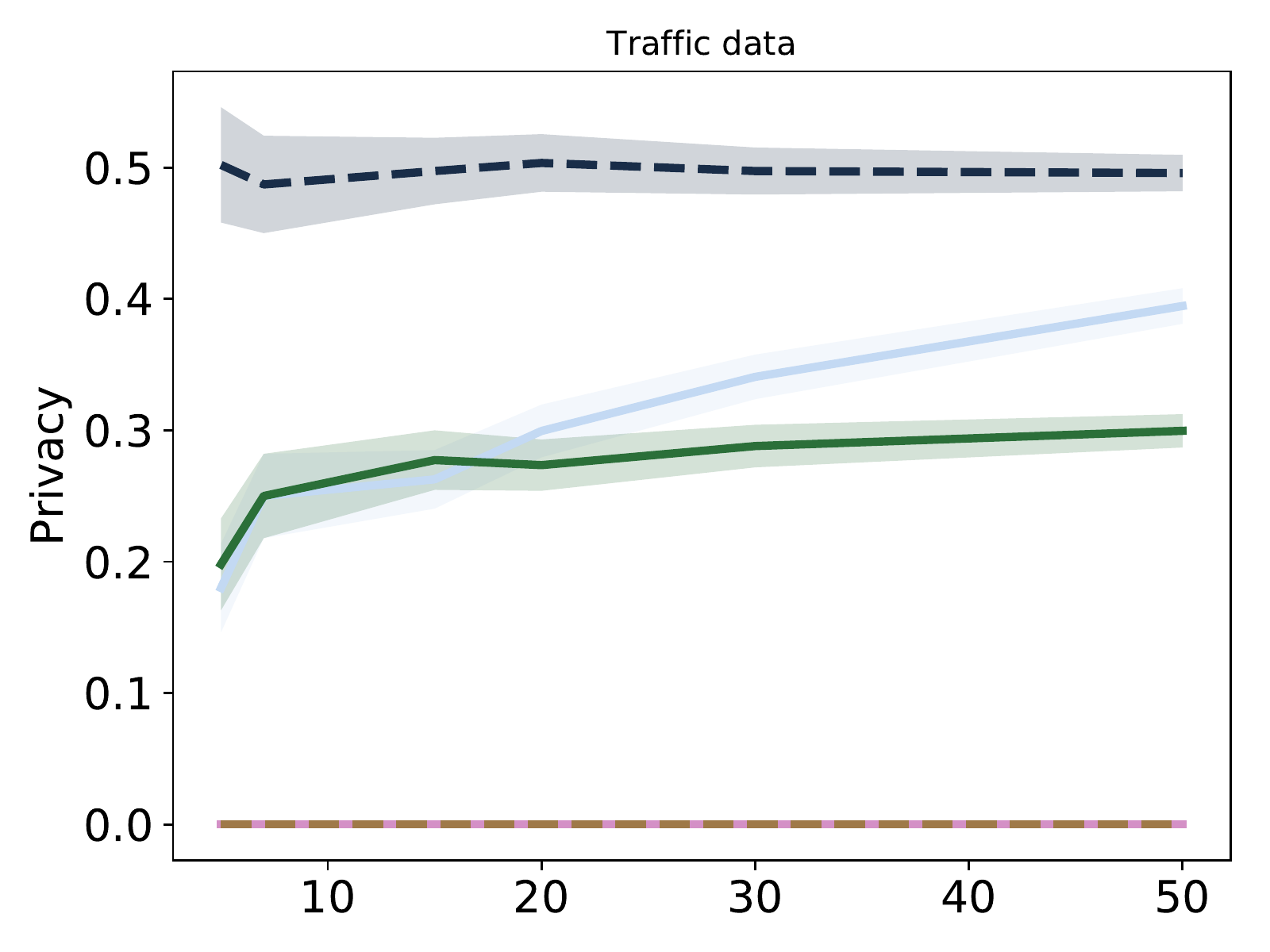}
      \includegraphics[width=\linewidth]{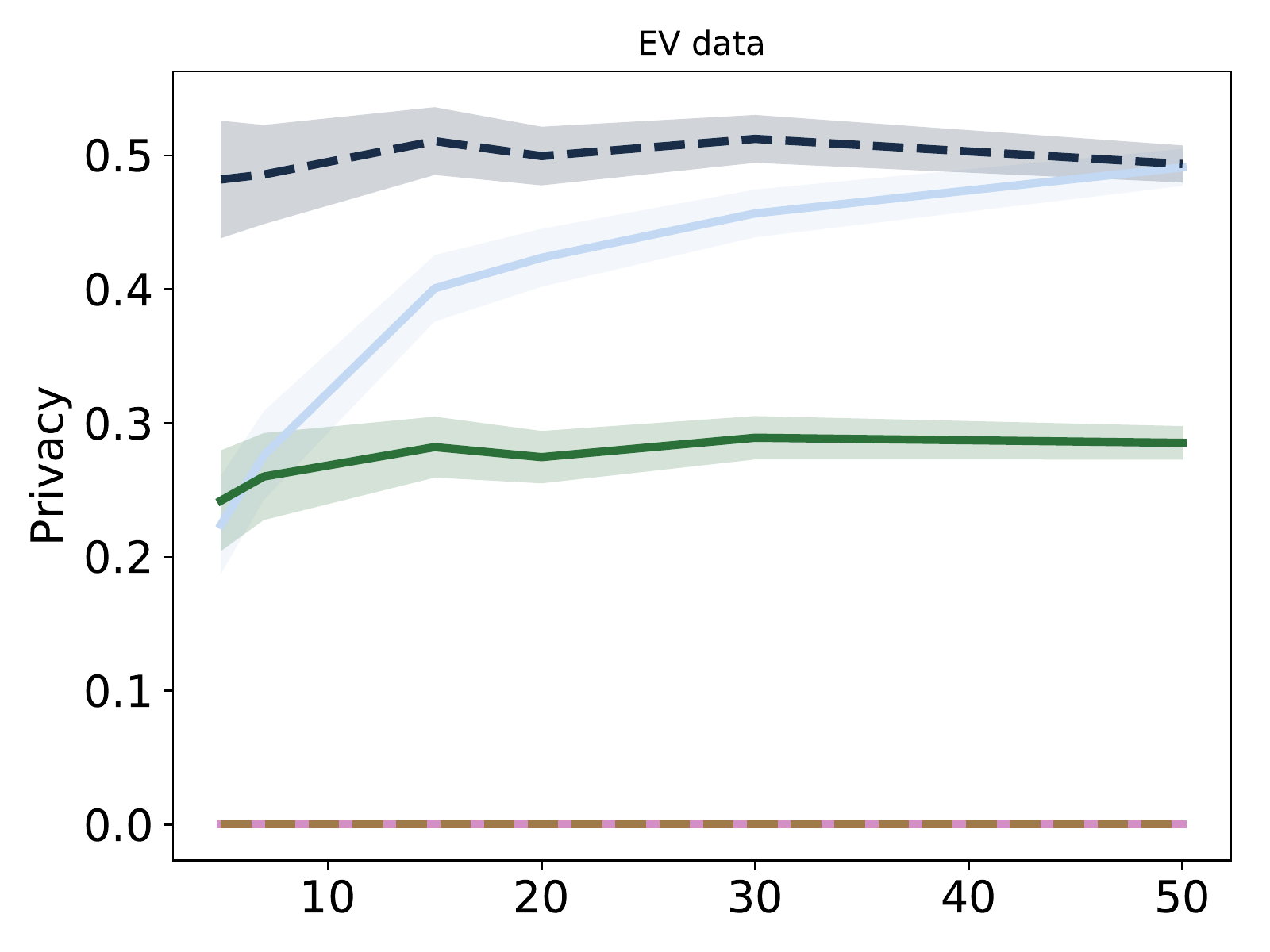}
        \subcaption{Average privacy of users.}\label{fig:priv}
  \end{minipage}%
  \begin{minipage}[b]{.33\linewidth}
    \includegraphics[width=\linewidth]{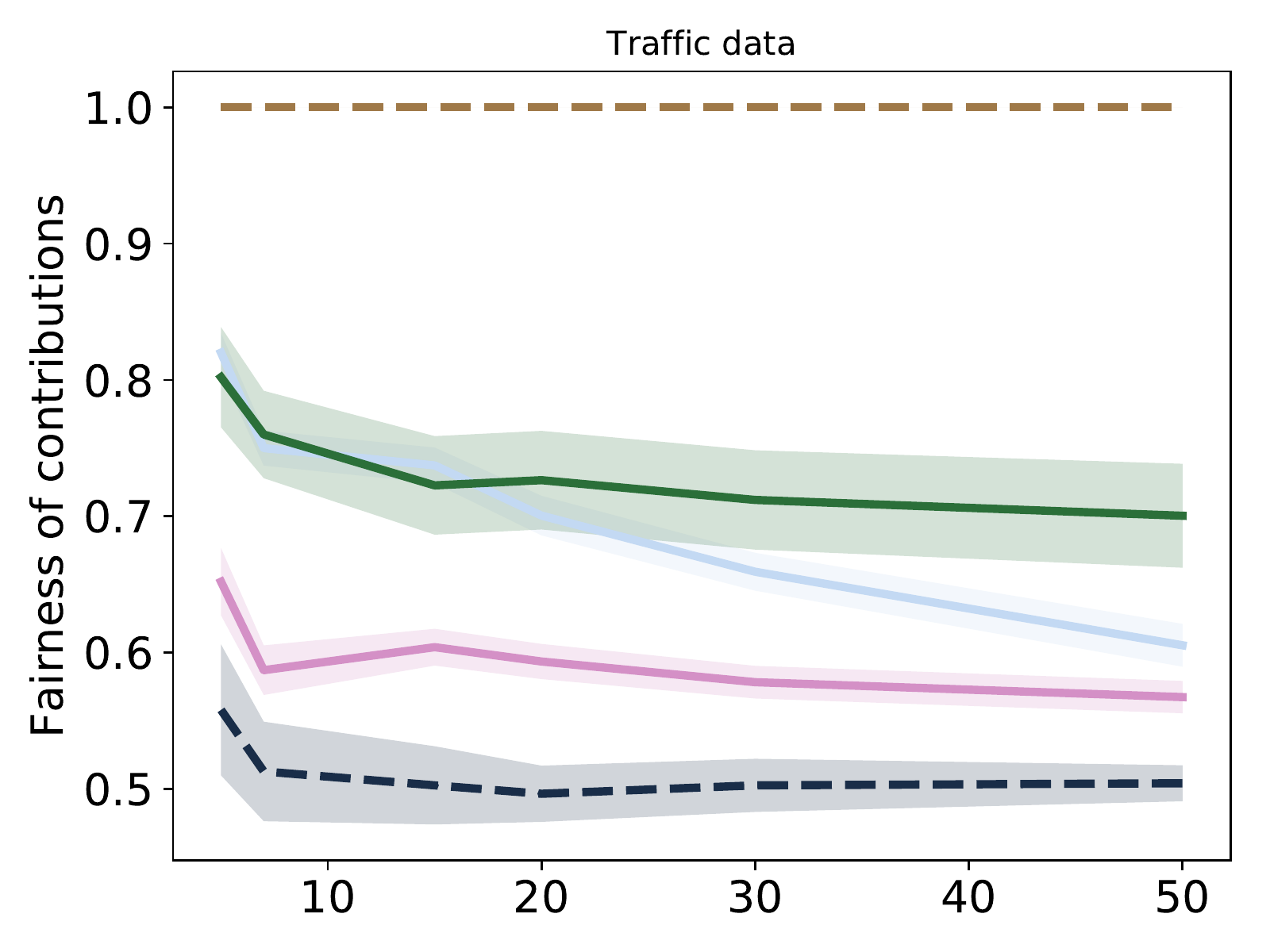}
    \includegraphics[width=\linewidth]{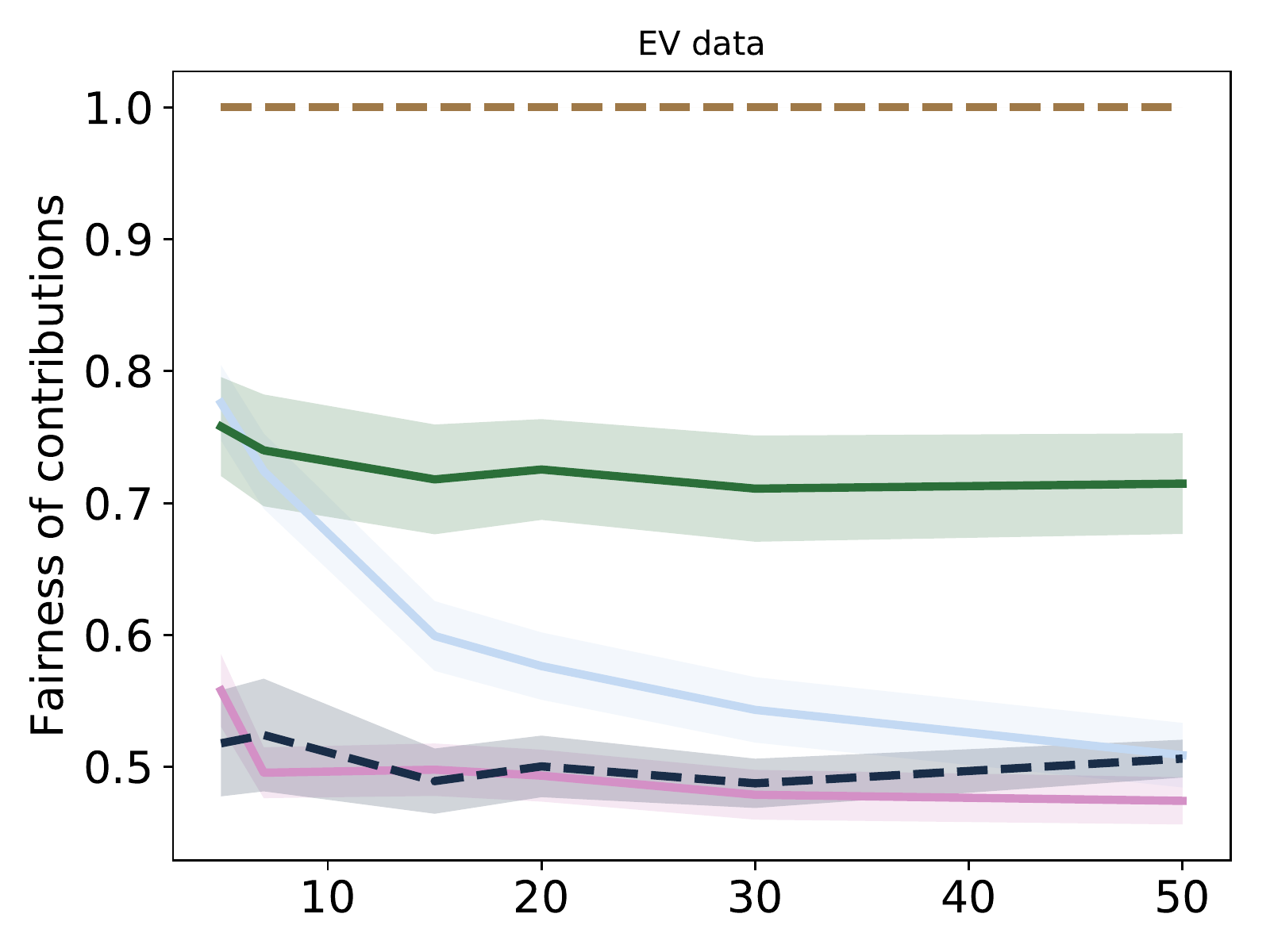}
        \subcaption{Average fairness of contributions.}\label{fig:gini}
  \end{minipage}%
  \begin{minipage}[b]{.33\linewidth}
    \includegraphics[width=\linewidth]{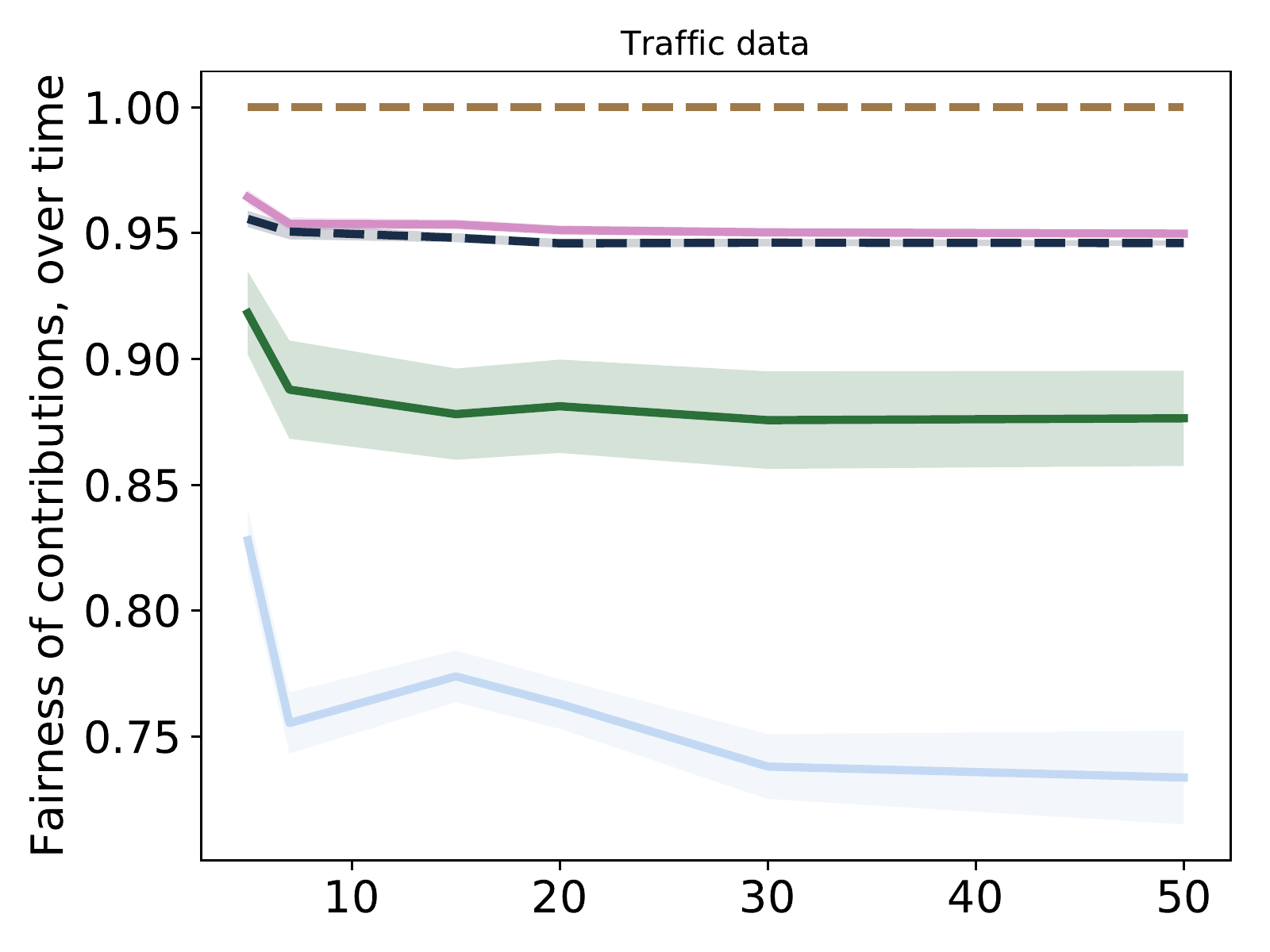}
    \includegraphics[width=\linewidth]{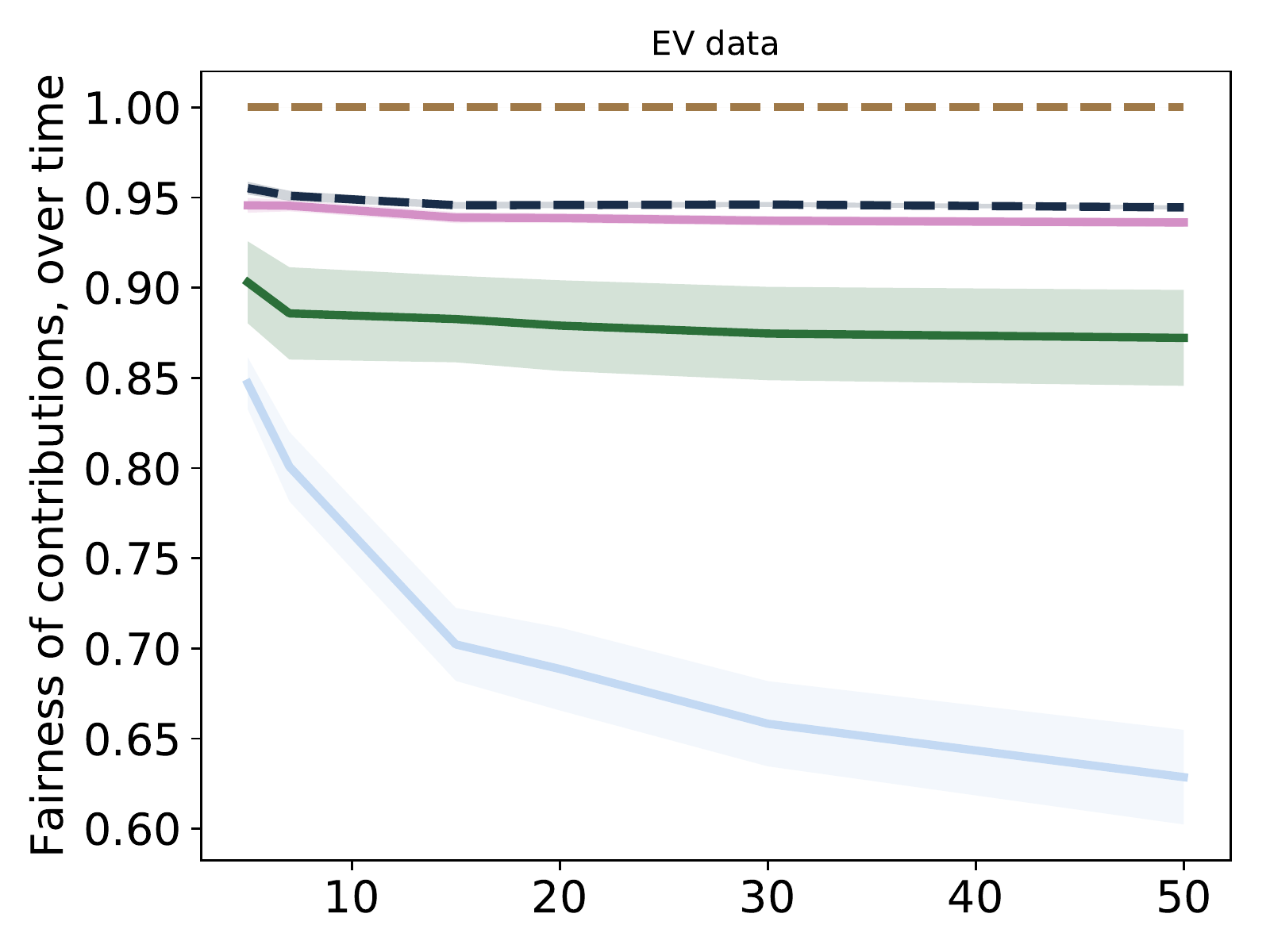}
        \subcaption{Average fairness over time.}\label{fig:ginihist}
  \end{minipage}
\end{minipage}
\caption{\textbf{Comparison of contribution strategies, the x axis represents the population size.} Dashed lines represent baselines, solid lines represent contribution strategies. The plots show trade-offs between centralized optimization, aspiration learning and Q-Learning in terms of efficiency, privacy and fairness. Optimization offers the highest success rate and efficiency, while aspiration learning and Q-Learning offers higher privacy. Q-Learning is the only contribution strategy to offer high fairness on both measures.}
\end{figure*}

In terms of success rate (Figure \ref{fig:succ}), full contribution and centralized optimization always succeed,i.e., are always able to provide the services, while Q-Learning does not guarantee success and fails in about 2-3\% of the cases.
In terms of efficiency (Figure \ref{fig:eff}), centralized optimization is the most efficient solution, as it finds the subset of users whose contribution satisfies the requirement \footnote{Optimization is not successful if the total requirement is larger than the sum of contributions from all agents, but the experiments are generated to be successful if all users contribute.} at the lowest cost \footnote{The approximation algorithm used to solve the optimization problem does not guarantee to find the global minimum.}.
Efficiency measures how close the total contribution approaches the needs of the system, hence baseline in which all agents contribute will have the lowest possible efficiency.
Efficiency increases with the size of the population as a higher number of possible solutions -- combinations of individual contributions -- makes it more likely to find an efficient solution.

Similarly, in terms of social welfare (see Figure \ref{fig:welfare}), optimization scores the highest, while localized strategies reach a welfare around 30\% lower.
Social welfare is the difference between the rewards from the public good and the costs of contributions, so a negative value indicates that costs are higher than gains.
Differently from the previous result, the performance of aspiration learning is equivalent to that of Q-Learning, as opposed to that of optimization.

In terms of privacy (Figure \ref{fig:priv}), centralized optimization achieves no privacy, like full contribution, as it requires knowledge about the state of all users.
Random contribution offers the highest privacy -- around 50\% of users -- at the expenses of other measures, e.g., success rate and efficiency.
Localized contribution strategies allow a fraction of the user to keep their data private.
This fraction increases with the population size for aspiration learning, while Q-Learning trades a lower privacy off for a higher fairness.

Fairness is measured in two ways: ``fairness of contributions'' compares the actions of all agents at the current time $t$ (see Figure \ref{fig:gini}), while ``fairness of contribution over time'' considers the histories of contribution up to time $t$ (see Figure \ref{fig:ginihist}).
Full contribution requires all users to contribute, thus it trades perfect fairness off for other measures such as efficiency.
Optimization offers low fairness because it considers only the current state, and users in certain states, e.g., with high values, are more likely to contribute than others.
Conversely, optimization offers high fairness over time because agents are randomly assigned to states, hence the chance of being in any state is over time the same.
This result might not hold if states are not randomly assigned, e.g., some users are more likely to obtain high values/costs than others.
Aspiration learning bases decisions only on the history of decisions, this leads to higher fairness, as contributions are independent of the state, and to lower fairness over time, caused by individual differences in training that accumulate over time.
These values decrease with the population size, while other contribution strategies are not affected by this parameter.
Q-Learning scores high values in both measures as it considers both the current context and the history of actions, this allows agents to learn similar behaviors by interacting with one another. 

\begin{table}[h]
  \centering{
  \begin{tabularx}{1.0\linewidth}{ll|ll}
    \textbf{Type} & \textbf{Algorithm} & \textbf{Pros} & \textbf{Cons} \\
    \hline
    Baseline & Full & Success & Efficiency \\
    Baseline & Random & Privacy & Success \\
    Centralized & Knapsack & Success, Efficiency & Privacy, Fairness \\
    Localized & Aspiration & Privacy & F. over time \\
    Localized & Q-Learning & Fairness (over time) & Efficiency \\
  \end{tabularx}
  \caption{Comparison of contribution strategies.}\label{tab:comparison}}
\end{table}

Summary of trade-offs is presented in Table \ref{tab:comparison}. Centralized optimization assures the success of the service and high efficiency, it is hence appropriate for mission critical services for which computation and network constraints are not an issue, e.g., measuring current load on the smart grid to prevent outages. Localized strategies are instead best suited for applications where privacy concerns might reduce user adoption, e.g., participatory sensing. Finally, Q-Learning offers the highest fairness, hence it is ideal for applications where fair access to the service is desirable, e.g., charging of electric vehicles.

\section{Conclusions and Future Work}
\label{sec:conclusions}

This paper provides a scenario-independent design principle and simulation framework for smart city applications that rely on voluntary user contribution.
Voluntary contributions empower users to control the ownership of their resource, e.g., by contributing data towards a service, independently of the type of resource and its use.
The applicability of this framework is verified using real-world data from the application scenarios of traffic congestion monitoring and electric vehicle charging.

Results quantify trade-offs produced by different contribution strategies along measures such as efficiency, privacy and fairness. The trade-offs identified hold in both scenarios, suggesting that they depend on characteristics of the algorithms and not on characteristics of the scenarios. Therefore the results can be used as implementation recommendations to service providers and system designers about the choice of contribution strategies.

Modeling and application in other scenarios with different cost and value characteristics is left to future work, for example ``negative'' contribution in traffic congestion, where users contribute to the public good by choosing a longer route instead of the shortest but congested route \cite{Huberman1997}. Incentives for contributions are important in scenarios that rely on user participation \cite{Radanovic2015}. However, incentives mechanisms would require quantifying privacy \cite{Norberg2007}, which could be addressed in future work. Another limitation of the current work is that only localized algorithms are considered. Relaxing this assumption is a worthy avenue for future work. Communication between agents would introduce privacy concerns during communication between agents, which would need to be measured, but would would allow analysis of new classes of algorithms, such as decentralized optimization.

\section{Acknowledgements}
Stefano Bennati acknowledges support by the European Commission through the ERC Advanced Investigator Grant ’Momentum’ [Grant No. 324247]

\end{document}